\documentstyle[twoside,epsf]{article}

\catcode`\@=11
\long\def\@makefntext#1{
\protect\noindent \hbox to 3.2pt {\hskip-.9pt  
$^{{\eightrm\@thefnmark}}$\hfil}#1\hfill}		%CAN BE USED 

\def\thefootnote{\fnsymbol{footnote}}
\def\@makefnmark{\hbox to 0pt{$^{\@thefnmark}$\hss}}	%ORIGINAL 
	
\def\ps@myheadings{\let\@mkboth\@gobbletwo
\def\@oddhead{\hbox{}
\rightmark\hfil\eightrm\thepage}   
\def\@oddfoot{}\def\@evenhead{\eightrm\thepage\hfil
\leftmark\hbox{}}\def\@evenfoot{}
\def\sectionmark##1{}\def\subsectionmark##1{}}

\oddsidemargin=\evensidemargin
\renewcommand{\thefootnote}{\fnsymbol{footnote}}

\newcounter{sectionc}\newcounter{subsectionc}\newcounter{subsubsectionc}
\renewcommand{\section}[1] {\vspace{12pt}\addtocounter{sectionc}{1} 
\setcounter{subsectionc}{0}\setcounter{subsubsectionc}{0}\noindent 
	{\tenbf\thesectionc. #1}\par\vspace{5pt}}
\renewcommand{\subsection}[1] {\vspace{12pt}\addtocounter{subsectionc}{1} 
	\setcounter{subsubsectionc}{0}\noindent 
	{\bf\thesectionc.\thesubsectionc. {\kern1pt \bfit #1}}\par\vspace{5pt}}
\renewcommand{\subsubsection}[1] {\vspace{12pt}\addtocounter{subsubsectionc}{1}
	\noindent{\tenrm\thesectionc.\thesubsectionc.\thesubsubsectionc.
	{\kern1pt \tenit #1}}\par\vspace{5pt}}
\newcommand{\nonumsection}[1] {\vspace{12pt}\noindent{\tenbf #1}
	\par\vspace{5pt}}

\newcounter{appendixc}
\newcounter{subappendixc}[appendixc]
\newcounter{subsubappendixc}[subappendixc]
\renewcommand{\thesubappendixc}{\Alph{appendixc}.\arabic{subappendixc}}
\renewcommand{\thesubsubappendixc}
	{\Alph{appendixc}.\arabic{subappendixc}.\arabic{subsubappendixc}}

\renewcommand{\appendix}[1] {\vspace{12pt}
        \refstepcounter{appendixc}
        \setcounter{figure}{0}
        \setcounter{table}{0}
        \setcounter{lemma}{0}
        \setcounter{theorem}{0}
        \setcounter{corollary}{0}
        \setcounter{definition}{0}
        \setcounter{equation}{0}
        \renewcommand{\thefigure}{\Alph{appendixc}.\arabic{figure}}
        \renewcommand{\thetable}{\Alph{appendixc}.\arabic{table}}
        \renewcommand{\theappendixc}{\Alph{appendixc}}
        \renewcommand{\thelemma}{\Alph{appendixc}.\arabic{lemma}}
        \renewcommand{\thetheorem}{\Alph{appendixc}.\arabic{theorem}}
        \renewcommand{\thedefinition}{\Alph{appendixc}.\arabic{definition}}
        \renewcommand{\thecorollary}{\Alph{appendixc}.\arabic{corollary}}
        \renewcommand{\theequation}{\Alph{appendixc}.\arabic{equation}}
        \noindent{\tenbf Appendix \theappendixc #1}\par\vspace{5pt}}
\newcommand{\subappendix}[1] {\vspace{12pt}
        \refstepcounter{subappendixc}
        \noindent{\bf Appendix \thesubappendixc. {\kern1pt \bfit #1}}
	\par\vspace{5pt}}
\newcommand{\subsubappendix}[1] {\vspace{12pt}
        \refstepcounter{subsubappendixc}
        \noindent{\rm Appendix \thesubsubappendixc. {\kern1pt \tenit #1}}
	\par\vspace{5pt}}

\newcommand{\textlineskip}{\baselineskip=13pt}
\newcommand{\smalllineskip}{\baselineskip=10pt}

\def\eightcirc{
\begin{picture}(0,0)
\put(4.4,1.8){\circle{6.5}}
\end{picture}}
\def\eightcopyright{\eightcirc\kern2.7pt\hbox{\eightrm c}} 

\newcommand{\copyrightheading}[1]
	{\vspace*{-2.5cm}\smalllineskip{\flushleft
	{\footnotesize International Journal of Modern Physics A, #1}\\
	{\footnotesize $\eightcopyright$\, World Scientific Publishing
	 Company}\\
	 }}

\def\abstracts#1#2#3{{
	\centering{\begin{minipage}{4.5in}\baselineskip=10pt\footnotesize
	\parindent=0pt #1\par 
	\parindent=15pt #2\par
	\parindent=15pt #3
	\end{minipage}}\par}} 

\newcommand{\bibit}{\nineit}

\renewenvironment{thebibliography}[1]
	{\frenchspacing
	 \ninerm\baselineskip=11pt
	 \begin{list}{\arabic{enumi}.}
	{\usecounter{enumi}\setlength{\parsep}{0pt}
	 \setlength{\leftmargin 12.7pt}{\rightmargin 0pt} %FOR 1--9 ITEMS
	 \setlength{\itemsep}{0pt} \settowidth
	{\labelwidth}{#1.}\sloppy}}{\end{list}}

\newcounter{itemlistc}
\newcounter{romanlistc}
\newcounter{alphlistc}
\newcounter{arabiclistc}

\newcommand{\fcaption}[1]{
        \refstepcounter{figure}
        \setbox\@tempboxa = \hbox{\footnotesize Fig.~\thefigure. #1}
        \ifdim \wd\@tempboxa > 5in
           {\begin{center}
        \parbox{5in}{\footnotesize\smalllineskip Fig.~\thefigure. #1}
            \end{center}}
        \else
             {\begin{center}
             {\footnotesize Fig.~\thefigure. #1}
              \end{center}}
        \fi}

\newcommand{\tcaption}[1]{
        \refstepcounter{table}
        \setbox\@tempboxa = \hbox{\footnotesize Table~\thetable. #1}
        \ifdim \wd\@tempboxa > 5in
           {\begin{center}
        \parbox{5in}{\footnotesize\smalllineskip Table~\thetable. #1}
            \end{center}}
        \else
             {\begin{center}
             {\footnotesize Table~\thetable. #1}
              \end{center}}
        \fi}

\def\@citex[#1]#2{\if@filesw\immediate\write\@auxout
	{\string\citation{#2}}\fi
\def\@citea{}\@cite{\@for\@citeb:=#2\do
	{\@citea\def\@citea{,}\@ifundefined
	{b@\@citeb}{{\bf ?}\@warning
	{Citation `\@citeb' on page \thepage \space undefined}}
	{\csname b@\@citeb\endcsname}}}{#1}}

\newif\if@cghi
\def\cite{\@cghitrue\@ifnextchar [{\@tempswatrue
	\@citex}{\@tempswafalse\@citex[]}}
\def\citelow{\@cghifalse\@ifnextchar [{\@tempswatrue
	\@citex}{\@tempswafalse\@citex[]}}
\def\@cite#1#2{{$\null^{#1}$\if@tempswa\typeout
	{IJCGA warning: optional citation argument 
	ignored: `#2'} \fi}}

\def\pmb#1{\setbox0=\hbox{#1}
	\kern-.025em\copy0\kern-\wd0
	\kern.05em\copy0\kern-\wd0
	\kern-.025em\raise.0433em\box0}

\def\fnt#1#2{\footnotetext{\kern-.3em
	{$^{\mbox{\scriptsize #1}}$}{#2}}}

\def\fpage#1{\begingroup
\voffset=.3in
\thispagestyle{empty}\begin{table}[b]\centerline{\footnotesize #1}
	\end{table}\endgroup}

\def\runninghead#1#2{\pagestyle{myheadings}
\markboth{{\protect\footnotesize\it{\quad #1}}\hfill}
{\hfill{\protect\footnotesize\it{#2\quad}}}}
\font\tenrm=cmr10
\font\tenit=cmti10 
\font\tenbf=cmbx10
\font\bfit=cmbxti10 at 10pt
\font\ninerm=cmr9
\font\nineit=cmti9

\font\eightrm=cmr8

\def\qed{\hbox{${\vcenter{\vbox{			%HOLLOW SQUARE
   \hrule height 0.4pt\hbox{\vrule width 0.4pt height 6pt
   \kern5pt\vrule width 0.4pt}\hrule height 0.4pt}}}$}}

\def\pt{$p^{}_T$}
\renewcommand{\thefootnote}{\fnsymbol{footnote}}	%USE SYMBOLIC FOOTNOTE

\begin{document}

\runninghead{A High-$P_T$ Trigger for the Hera-B Experiment}
{A High-$P_T$ Trigger for the Hera-B Experiment}

\normalsize\textlineskip
\thispagestyle{empty}
\setcounter{page}{1}

\copyrightheading{}			%{Vol. 0, No. 0 (1993) 000--000}

\vspace*{0.88truein}

\fpage{1}
\vspace*{-0.5in}
\begin{flushright}
{\normalsize UCTP-115-00}
\end{flushright}
\vskip0.20in
\centerline{\bf A HIGH-$P_T$ TRIGGER FOR THE HERA-B EXPERIMENT}
\vspace*{0.37truein}
\centerline{\footnotesize A. K. MOHAPATRA\footnote{e-mail : 
mohapatr@mail.desy.de}} 
\vspace*{0.015truein}
\centerline{\footnotesize\it Department of Physics, University
of Cincinnati}
\baselineskip=10pt
\centerline{\footnotesize\it Cincinnati, OH 45221-0011, USA}
\vspace*{7pt}
\vspace*{0.225truein}
%\publisher{(received date)}{(revised date)}
\vspace*{8pt}

\vspace*{0.21truein}
\abstracts{We have constructed a high-$p_T$ trigger for the HERA-B
experiment at DESY. The HERA-B experiment produces B mesons by
inserting wire targets into the halo of the proton beam circulating in
HERA. The high-\pt\ trigger records events that contain tracks that have
high transverse momentum with respect to the beam. Such a trigger is
efficient for recording $B \rightarrow \pi^+\pi^-$, $B \rightarrow
K^-\pi^+$, $B_s \rightarrow K^+ K^-$, $B_s \rightarrow D_s^-\pi^+$,
and other topical hadronic B decays. These decays provide sensitivity
to the internal angles $\alpha$ and $\gamma$ of the CKM unitarity triangle,
and they also can be used to measure or constrain the $B_s$-$\bar B_s$ 
mixing parameter $x_s$.}{}{}

\textheight=7.8truein
\setcounter{footnote}{0}
\renewcommand{\thefootnote}{\alph{footnote}}

\vspace*{4pt}
\vspace*{1pt}\textlineskip      %) USE THIS MEASUREMENT WHEN THERE IS
\section{Introduction}          %) A SECTION HEADING
\vspace*{-0.5pt}
\noindent
The construction of a high-$p_T$ trigger is an attempt to broaden 
the physics reach of the HERA-B experiment beyond it's main goal
of measuring sin$2\beta$ through the decay mode 
$B^0\rightarrow J/\psi K^0_s$. Some of the possibilities include 
measurement of the angles $\alpha$ and $\gamma$ through hadronic B
decays such as $B^0\rightarrow\pi^+ \pi^-,\ \pi^- K^+$ and 
$B^0\rightarrow \pi^\pm K^\mp,\ D^{*+} \pi^-$, respectively.\cite{PHYS} 
In addition, this trigger can be useful to constrain the $B_s$-$\bar B_s$
mixing parameter $x_s$.

The HERA-B experiment produces $b$ mesons and baryons by colliding 920
GeV/c protons in the HERA storage ring with 4-8 fixed wire
targets. The nominal interaction rate is 20-40 MHz. The configuration of
detectors downstream of the targets are: a silicon vertex detector
for reconstruction of decay vertices; a tracking system for momentum
measurement; and lepton ($e$, $\mu$) and hadron ($\pi$, $K$ and $p$)
identification systems consisting of an electromagnetic calorimeter, 
gas ionization chambers interleaved with iron plates, and a Ring-Imaging
Cherenkov counter (RICH). Detailed descriptions of these
detectors can be found elsewhere.\cite{TDR}

\vspace*{1pt}\textlineskip             
\section{High-$P_T$ Trigger System}    
\vspace*{-0.5pt}
\noindent
The objective of the high-\pt\ trigger system is to identify
high-\pt\ hadrons. This is done by selecting high-\pt\ track
candidates in an event based on the hit patterns in the trigger
chambers with a pad-type readout placed within the spectrometer
magnet. The hit positions determine ``regions-of-interest'' in downstream
tracking chambers which are used as seeds for a Kalman-filter tracking
algorithm in the First Level Trigger (FLT).

\begin{figure} [t]
\vspace*{13pt}
\centerline{
\epsfxsize=5cm\epsfbox{ 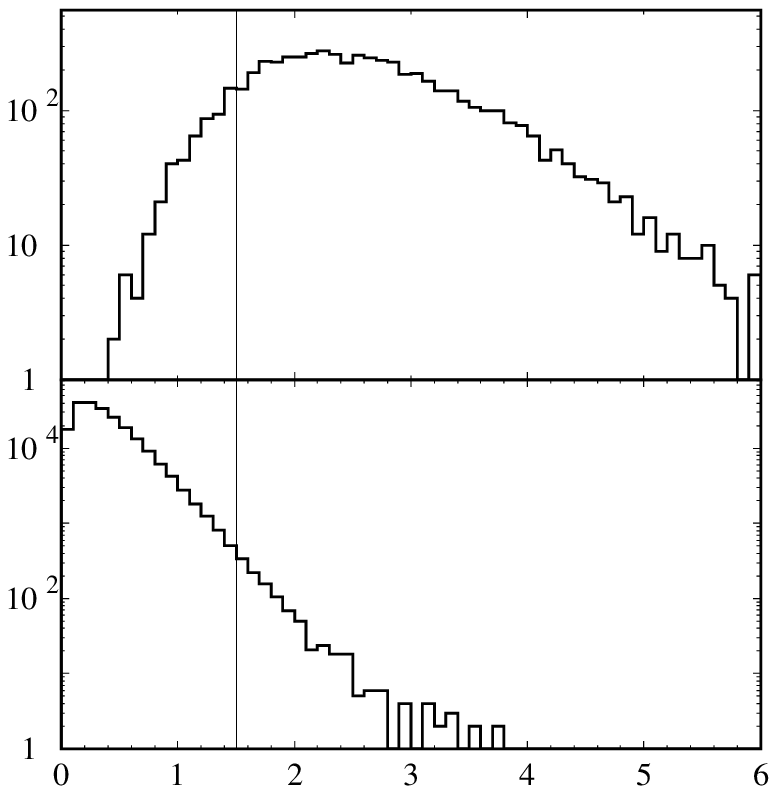}}
\vspace*{13pt}
\fcaption{{\footnotesize \it Momentum (GeV/c) distribution of pions from
           $B^0\rightarrow \pi^+ \pi^-$ decays (upper plot) and of 
	   tracks from minimum bias events (lower plot). The vertical 
	   scale is (N/0.02) GeV/c for both plots.}}
\label{ptspec}
\end{figure}

The high-\pt\ tracks typically bend little or make larger angles with
respect to the beam direction, and their hit patterns in the trigger 
chambers are distinct from those caused by minimum bias tracks. The 
\pt\ distribution for tracks from two body decays of $b\bar b$ and 
minimum bias events are shown in Fig.~\ref{ptspec}. This difference 
in the \pt\ spectrum has been exploited in the design of this trigger
system.

The high-\pt\ trigger chambers (12 in total) with pad readout are
organized as follows: six gas-pixel chambers\cite{PIXEL} for the
region nearest the beam, and six straw-tube chambers with pad 
readout\cite{STRAW} for the larger outer region. Half the chambers are
located on the $+x$ side of the beam and half on the $-x$ side. For each
side, the six chambers are positioned at three different distances from
the target (along the beam direction); these are referred to as the 
PT1, PT2, and PT3 stations. The inner chambers have a total of 11960 
channels and an angular acceptance in the bending view of 
10\,$<$\,$\theta$\,$<$\,60~mrad. The outer chambers
have a total of 6240 channels and an angular acceptance of 
40\,$<$\,$\theta$\,$<$\,250~mrad. The gas-pixel chambers 
were constructed at ITEP, Moscow. The outer straw tube 
chambers were constructed at the University of Cincinnati 
and Princeton University.

\vspace*{1pt}\textlineskip             
\section{Implementation of the Trigger Scheme}
\begin{figure} [htpb]
\centerline{
\epsfxsize=6cm\epsfbox{ 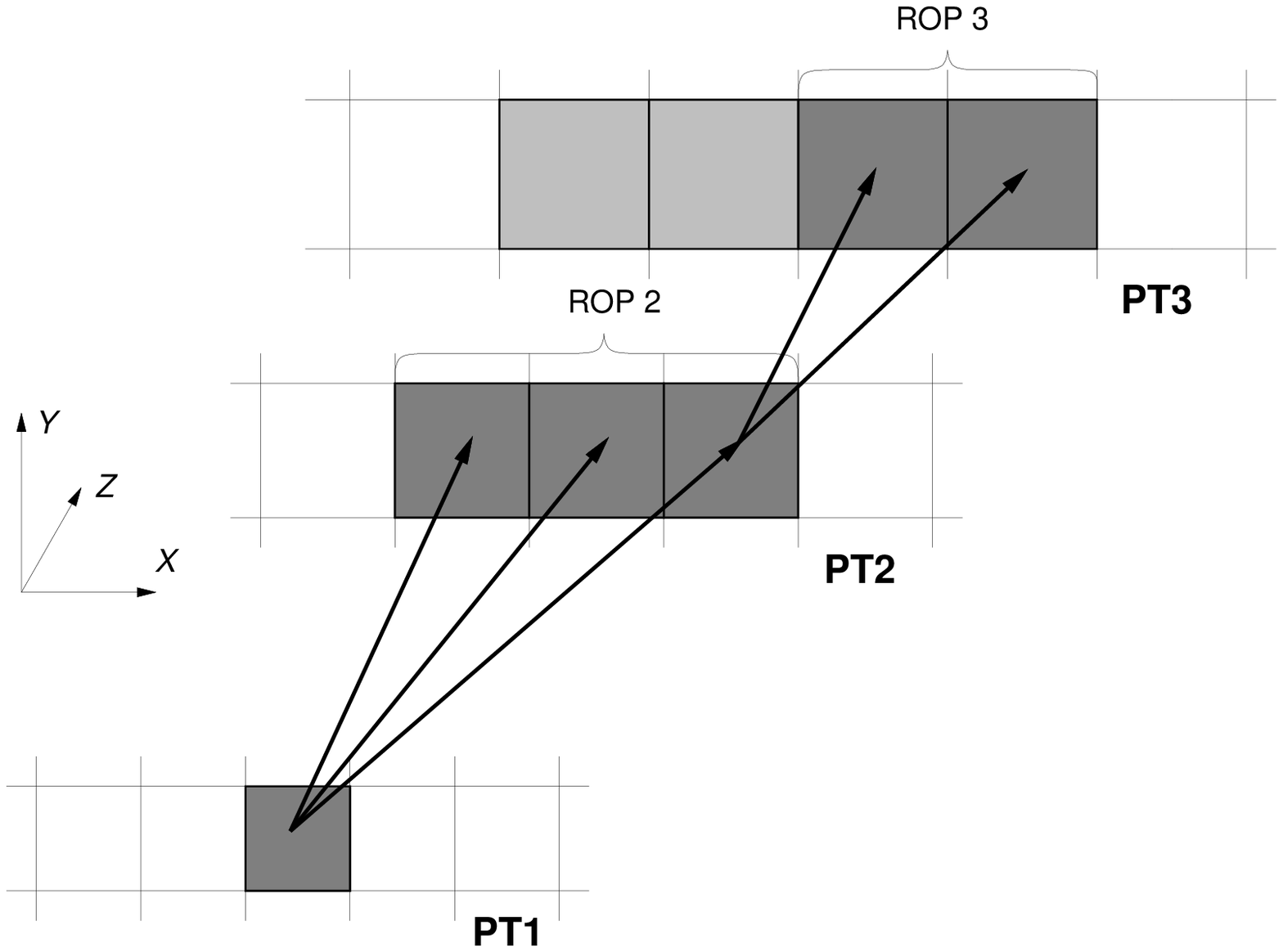}}
\vspace*{13pt}
\fcaption{{\footnotesize\it Sketch of the pad coincidence algorithm
            for the PT1, PT2, and PT3 chambers.}} 
\label{coin}
\end{figure}

\begin{figure} [htbp]
\vspace*{13pt}
\centerline{
\epsfxsize=5cm\epsfbox{ 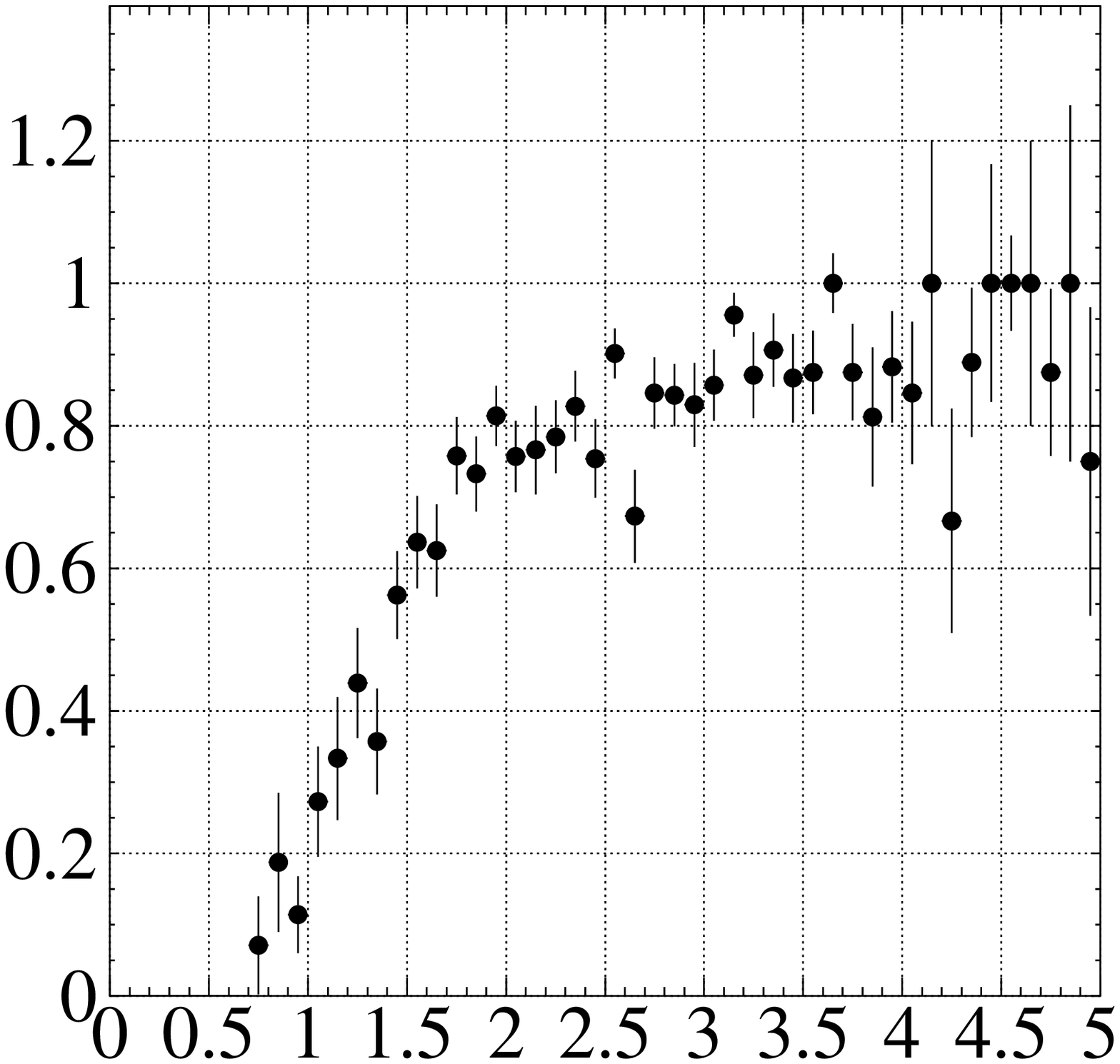}}
\vspace*{13pt}
\fcaption{{\footnotesize\it The trigger efficiency versus the \pt\ (GeV/c) of
           the track}.}
\label{eff}
\end{figure}

\vspace*{-0.5pt}
\noindent
The nominal high-\pt\ track pre-selection is based on a 1-3-2
coincidence algorithm: combination of one pad in PT1 with three pads
in PT2, and each of the three pads in PT2 with two pads in PT3. 
A sketch of this triple-pad-coincidence scheme is shown in
Fig.~\ref{coin}.  The scheme is implemented in the trigger electronics 
and can be modified in situ depending on the background level.
Based on the 1-3-2 algorithm, the Monte Carlo estimation of trigger
efficiency for pions from $B \rightarrow \pi^+\pi^-$ decays is
shown in Fig.~\ref{eff}. With an input data rate of $2 \times 10^{11}$
bits per second, the high-\pt\ trigger system is expected to process
approximately $3 \times 10^{12}$ coincidence patterns (for 1-3-2
algorithm) per second and send the kinematic information to the
FLT. At an interaction rate of 40 MHz, the typical output data rate is
expected to be ${\cal O}(10^7)$ messages per second. The high-\pt\ 
trigger logic hardware is designed to successfully handle 
$1.5 \times 10^8$ messages per second.

\nonumsection{References}

\end{document}